\documentclass[conference,a4paper]{IEEEtran}
\usepackage{amsmath}
\usepackage{amssymb}
\usepackage{algorithmic}
\usepackage{array}
\usepackage{cite}
\usepackage{graphicx}
\usepackage{hyperref}
\usepackage{url}
\usepackage{siunitx}
\usepackage{tikz}

\DeclareSIUnit{\dBm}{dBm}
\newcommand\copyrighttext{%
\footnotesize \textcopyright \enspace 2016 IEEE. Personal use of this material is permitted. Permission from IEEE must be obtained for all other uses, in any current or future media, including reprinting/republishing this material for advertising or promotional purposes, creating new collective works, for resale or redistribution to servers or lists, or reuse of any copyrighted component of this work in other works.  DOI: \href{https://doi.org/10.1109/WoWMoM.2016.7523579}{10.1109/WoWMoM.2016.7523579}
}
\newcommand\copyrightnotice{%
\begin{tikzpicture}[remember picture,overlay]
\node[anchor=south] at (current page.south) {\fbox{\parbox{\dimexpr\textwidth-\fboxsep-\fboxrule\relax}{\copyrighttext}}};
\end{tikzpicture}%
}

\begin{document}
\title{Beacons in Dense Wi-Fi Networks: \\ How to Befriend with  Neighbors in the 5G World?}
\author{Dmitry Bankov\IEEEauthorrefmark{1}, Evgeny Khorov\IEEEauthorrefmark{1}, Andrey Lyakhov\IEEEauthorrefmark{1}, Sigurd Schelstraete\IEEEauthorrefmark{4}\\\IEEEauthorrefmark{1}Institute for Information Transmission Problems, Russian Academy of Sciences, Moscow, Russia\\\IEEEauthorrefmark{4}Quantenna Communications, US\\Email: \{bankov, khorov, lyakhov\}@iitp.ru, sschelstraete@quantenna.com 
}
\maketitle
\copyrightnotice

\begin{abstract}
To address 5G challenges, IEEE 802.11 is currently developing new amendments to the Wi-Fi standard, the most promising of which is 802.11ax. 
A key scenario considered by the developers of this amendment is dense and overlapped networks typically present in residential buildings, offices, airports, stadiums, and other places of a modern city. 
Being crucial for Wi-Fi hotspots, the hidden station problem becomes even more challenging for dense and overlapped networks, where even access points (APs) can be hidden.
In this case, user stations can experience continuous collisions of beacons sent by different APs, which can cause disassociation and break Internet access. 
In this paper, we show that beacon collisions are rather typical for residential networks and may lead to unexpected and irreproducible malfunction. We investigate how often beacon collisions occur, and describe a number of mechanisms which can be used to avoid beacon collisions in dense deployment.  
Specifically, we pay much attention to those mechanisms which are currently under consideration of the IEEE 802.11ax group. 
\end{abstract}

\section{Introduction}
\label{sec:intro} 

Today it is apparent that to meet heterogeneous or even incompatible 5G requirements, the 5G architecture shall look like a system of systems \cite{nokia_system_of_systems}, joining several advanced technologies which complement each other, even though a lot of effort is being put to improve the technologies both on the PHY \cite{zhilin2015high} and MAC layers.
Specifically, industry leaders and network researchers consider 5G as a synergy of next generation LTE and Wi-Fi as well as a number of novel technologies.
Although some LTE specialists have a skeptical view of Wi-Fi, which works in unlicensed spectrum, few would dispute that it has become a must, and it is everywhere: in offices, homes, restaurants, airports, parks.
The number of Wi-Fi-enabled devices is continuously growing as well as the number of Wi-Fi hotspots.
Even today Wi-Fi can provide multigigabit Internet access, and cellular networks are incapable of providing such high rates. 

To address 5G challenges and to improve Wi-Fi performance in a vast range of scenarios relevant to 5G networks, the IEEE 802 LAN/MAN Standard Committee has launched the High Efficiency Wireless local area networks (WLAN) Study Group (HEW SG), which was later transformed into the Task Group AX (TGax) or IEEE 802.11ax.
By 2019, the group aims to develop a new amendment to the Wi-Fi standard, which will improve performance of Wi-Fi networks in challenging existing and emerging scenarios.
An important issue considered by the group is Wi-Fi efficiency in dense deployment\cite{etnt}.

The ubiquity of Wi-Fi and the need of higher throughput results in densification: the more access points (APs) you have, the better is the coverage and the higher are the data rates.
The reverse side of the medal is dozens of available APs not only in public places, but even in homes \cite{chuck}.
A high number of devices working at the same area and in the same channel results in collisions.
Based on random channel access Wi-Fi
suffers from dense deployment. However the more crucial for
Wi-Fi networks is the hidden station problem, which in case
of overlapped networks becomes even more challenging, since
even APs can be hidden.

Being hidden from each other, the APs can transmit in overlapping time intervals.
In case of data transmission, it results just in collisions, retries and inefficient channel usage.
In addition to data, APs transmit beacons, which are the most important management frames in the network.
Beacons are transmitted strictly periodically and contain vital information for synchronization and network operation.
Apart from that, they signal AP presence, and if several beacons are lost in a row, a client station (STA) can assume that it has gone out of the transmission range of the AP and disassociate from it.
Since beacons are transmitted periodically, and almost nobody changes the default beacon interval, if beacons of two APs start to collide, they will collide for ages, until clock drifting or an AP reboot correct the situation.

In this paper, we show that the probability of beacon collisions is quite high in residential networks and may lead to unexpected and irreproducible malfunction.
We study the frequency of beacon collisions, and propose a way to avoid them in dense deployment, paying much attention to approaches that are currently (March, 2016) under consideration of the IEEE 802.11ax group.
The rest of the paper is organized as follows.
Section \ref{sec:related} describes the current state of the research on beacon collisions.
Section \ref{sec:scenario} describes the studied scenario.
In Section \ref{sec:model} we present the used notation and list the conditions in which the beacon collisions occur.
Section \ref{sec:results} contains the experimental results and the discussion about the probability of beacon collisions and the ways to avoid them.
Section \ref{sec:conclusion} concludes the paper.

\section{Background}
\label{sec:related}

Early research on beacon collisions in Wi-Fi networks dates back to the year 2000. For example, the authors of~\cite{cervello2000beacon} show that the performance of the Point Coordination Function (PCF) --- a contention free channel access mechanism --- degrades in case of beacon collisions. Specifically, the authors consider a single Wi-Fi hot spot where beacons from an AP collide with data frames transmitted by the STAs associated with the AP. Their results show that the probability that a beacon collides with a data frame can reach 4\% for five STAs in the network.

Although such occasional and rare beacon collisions can degrade the network performance, e.g. by affecting power management and time synchronization mechanisms, the consequences are hardly dramatic.

The more crucial is the problem of beacon-to-beacon collisions, which was carefully studied for wireless multihop networks \cite{lee11, lee2013} and vehicular networks\cite{Vinel}. This area is of high importance since the studies have finished with a number of solutions which can be favorable for dense networks.

In Wi-Fi Mesh networks specified in \cite{802.11}, beacons are broadcast by every STA to notify the presence of the STA in the neighborhood.
Consider three STAs: A, B, and C, such that pairs of STAs (A and B) and (B and C) are in the transmission range of each other, respectively, while A and C are hidden.
Since beacons are transmitted strictly periodically with the same period, having started to collide, beacons of A and C continue colliding for ages.
Since B receives beacons neither from A nor from C, it will never get information about these STAs. 

This problem has been studied in \cite{vishnevsky2007beaconing} in a multi-hop scenario, with STA layout similar to the one studied in our paper.
The studied case presents a mesh network in a three-level building with three rooms on a floor.
Every room contains three mesh STAs and the corridors and stairs contain additional STAs to forward the traffic to the gateway.
The authors have studied the time that a multi-hop path can survive, the path being destroyed when one of the STAs on the path cannot transmit its beacon for a specific time interval.
They have shown that with standard beacon transmission approach the probability of successful beacon transmission is less than 30\%.
In order to deal with this problem the authors propose a p-persistent approach the idea of which is that each beacon interval a STA attempts to transmit its beacon with the given probability $p$.
As the result, beacon collisions occur more rarely and a significant increase in path survival time is observed.

The standard \cite{802.11} proposes two ways how to avoid beacon collisions (Mesh Beacon Collision Avoidance).
First, each mesh STA advertises beacon transmission times of its neighbors and uses the received information to adjust its own beacon transmission time.
Second, beacons can be delayed by a random time to discover beacon collisions.
Note that deferring transmission by a random time aka jitter is a rather common approach widely used in many neighborhood discovery and routing protocols developed for Mobile Ad hoc Networks (MANETs) by IETF \cite{nhdp, olsr, kureev}.

Dense network deployment differs from the mesh network (or MANETs) in several aspects, which raises the question if beacon collisions are as probable and as crucial in dense networks as in the mesh networks.
First, in mesh networks there are long links with low signal to noise ratio (SNR), while in typical scenarios for dense infrastructure networks the STAs are located close to the AP to which they are connected.
Second, the set of operating channels for neighboring mesh STA shall overlap, to allow direct transmission between them.
In contrast, to reduce inter-network interference for dense deployment, it is more typical to use different channels.
For these reasons, when the beacon collision problem has been discussed during the recent January IEEE 802.11 meeting, the reaction of the group was that the problem of beacon collisions in dense networks, the impact of beacon collisions on the network performance and ways to avoid the beacon collisions shall be studied in the framework of the 802.11ax scenarios of interest. 

In this paper, we present the results of such a study, focusing on the key 802.11ax residential scenario~\cite{presentation_scenarios}.      

\section{Scenario}
\label{sec:scenario}

\begin{figure}[htb]
	\centering
	\includegraphics[width=0.9\linewidth]{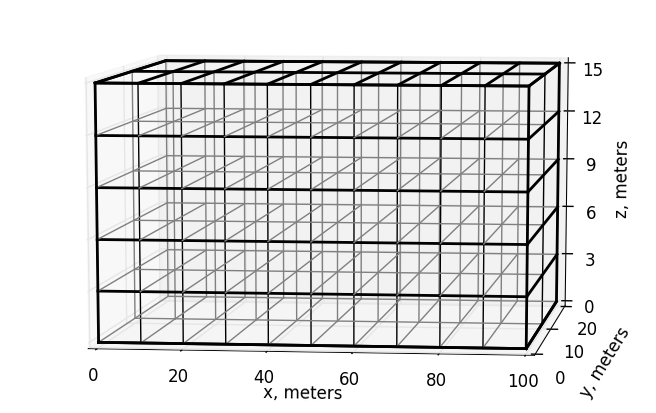}
	\caption{\label{fig:layout} Residential building}
\end{figure}

In the residential scenario, we consider a 5-floor building.
At every floor, we have two rows of apartments with 10 apartments of equal size in each row (see Fig.~\ref{fig:layout}).
There are an AP and several STAs in each apartment. 

Since 802.11ax extends 802.11ac PHY and support both 2.4 and 5 GHz bands, we consider two cases, mentioned in~\cite{presentation_scenarios}.
In the first case, all APs work in a \SI{2.4}{\GHz} band and every AP randomly selects one of three available orthogonal \SI{20}{\MHz} channels, which corresponds to 802.11b/g/n.
In the second case, all APs work in a \SI{5}{\GHz} channel, which corresponds to 802.11ac.
Each AP randomly chooses one of 3 or 5 orthogonal \SI{80}{\MHz} channels.
Then it randomly selects the primary \SI{20}{\MHz} channel, i.e. the channel where beacons are sent.
Thus, if we use \SI{5}{\GHz}, the total number of primary channels is 12 or 20, respectively. 

We also assume that every AP transmits a beacon of duration \SI{0.5}{\ms} with period of \SI{0.5}{\s}.
No other traffic is transmitted.

\section{Methodology}
\label{sec:model}

Let us analyze how often collisions of beacons sent by hidden APs occur.
Specifically, we are interested in probability of such collisions that prevent a STA from obtaining a beacon from the AP located in the same apartment.
Such a beacon collision happens when the following three conditions hold: 
\begin{itemize}
	\item \textbf{Time Condition}:  the beacons overlap in time, 
	\item \textbf{Channel Condition}: they are sent in the same channel,
	\item \textbf{Location Condition}: the APs that send the beacon and the STA at which the collision occurs are located in a specific way described below. 
\end{itemize} 

\subsection{Time Condition}

At the first sight, a beacon collision always occurs when beacons overlap in time.
However, let us consider an apartment and assume, that the signal strength from the AP located in this apartment (denoted as $AP_0$) is much stronger than the signal from the AP from another apartment ($AP_x$).
Note that although such an assumption is plausible, it does not always hold. 

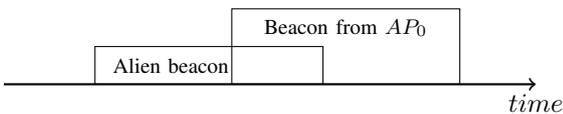
\begin{figure}[htb]
	\centering
	\begin{tikzpicture}
		\draw[thick,->] (0,0) -- (7,0) node[below] {$time$};
	\draw (3,0) rectangle (6,1);
	\draw (1.2,0) rectangle (4.2,0.5);
	\draw (4.5,0.75) node {\footnotesize{Beacon from $AP_0$}};
	\draw (2.2,0.25) node {\footnotesize{Alien beacon}};
	\end{tikzpicture}
	\caption{\label{fig:tc1} Time condition}
\end{figure}

Under such an assumption, if a beacon from $AP_0$ starts earlier, an alien beacon from $AP_x$ is just a weak noise and cannot damage the beacon from $AP_0$.
In contrast, if the alien beacon starts earlier, the STA starts receiving this beacon and does not sync at the beacon from $AP_0$%
\footnote{This statement needs some explanation.

In many sources \cite{choi2005performance, leentvaar1976capture}, the capture effect is described. Specifically, this effect means that a high-power frame begins while a STA is receiving a low-power frame, the STA stops receiving the old frame and resynchronizes to the new one. In Wi-Fi networks, capture effect does present but only when the STA receives the frame preamble, which lasts for dozens of microseconds, while the whole beacon is much longer. After that, the STA receives the frame length and, according to the standard receive state machine, does not start to look for the preamble until the frame ends, even if the synchronization is lost due to the noise or an intersecting frame that is transmitted with more powei. Although looking for a new preamble during a frame reception is implemented by some vendors, it requires to complicate devices and can degrade performance in some cases. Thus in this paper, we do not take capture effect into account.
}.
So, we can consider a stricter time condition. This condition is met when an alien AP sends a beacon that starts earlier than a beacon of  the AP to which a STA is associated, and intersects this beacon (see Fig.~\ref{fig:tc1}).

Since beacons are transmitted periodically, the probability that the time condition is met for one alien AP is $l/B$, where $l$ is the beacon duration (except for the preamble duration) and $B$ is the beacon interval.

\subsection{Channel Condition}

The Channel Condition is met when the APs send beacons in the same channel.
Since the channels are selected randomly, the probability that two APs have the same primary \SI{20}{\MHz} is just $1/N$, where $N$ is the number of available orthogonal \SI{20}{\MHz} channels.
%

%
%
%

\subsection{Location Condition}

\begin{figure}
\centering
\includegraphics[width=0.4\linewidth]{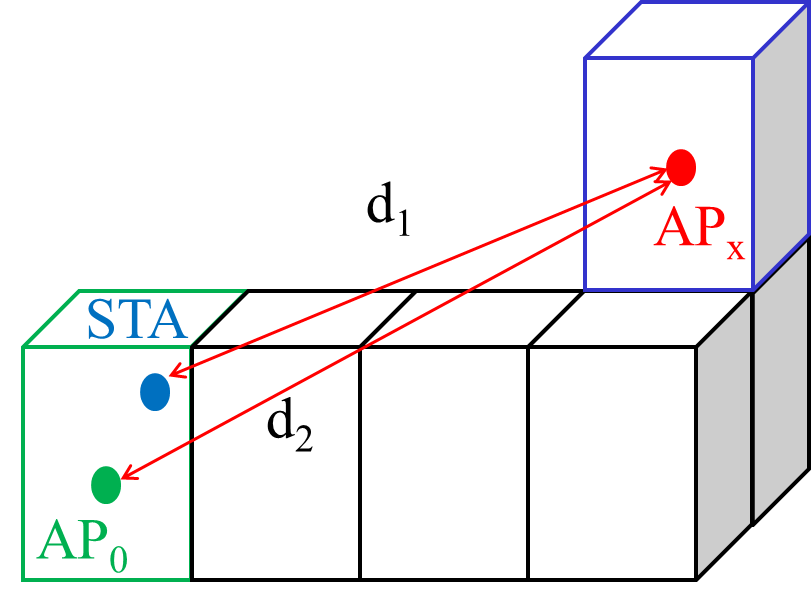}
\caption[Location condition]{Location condition}
\label{fig:location}
\end{figure}

Consider a STA, $AP_0$ located in the same apartment, and $AP_x$ located in another apartment, see Fig.~\ref{fig:location}.
A beacon collision happens at the STA, if it receives the beacon from $AP_x$, while $AP_0$ does not sense this beacon and starts to transmit its own one.
$AP_0$ does not sense the beacon from $AP_x$, if the strength of the signal received at $AP_0$ is less than the sensitivity threshold $P_{th}$ \cite{presentation_scenarios}.
At the same time, the STA receives the signal, if its strength is above this threshold.
To take into account the fading and the fact that APs typically have higher antenna gain than STAs, we introduce parameter $\Delta P$ and further consider a stricter location condition:

\begin{equation}
\label{eq:loc}
\begin{cases}
P(STA, AP_x) \geq P_{th} + \Delta P,\\
P(AP_{0}, AP_x) < P_{th},
\end{cases}
\end{equation} 
where $P(A,B)$ is the power of the signal from device $B$ received at device $A$. 

For the residential scenario, the pass loss model can be determined by the following equation~\cite{presentation_scenarios}:
\begin{align*}
PL(d) &= 40.05 + 20 log_{10}(\frac{f_c}{2.4}) + 20 log_{10}(min(d, 5)) + \\
&+ \mathbb{I}(d > 5) 35 log_{10}(\frac{d}{5}) + 18.3 F^{\frac{F + 2}{F + 1} - 0.46} + 5 W,
\end{align*}
where $PL$ is the signal loss expressed in dB, $d$ is the distance between devices [\SI{}{\m}] (note that the equation is valid only for $d>1$),  $f_c$ is the central channel frequency [\SI{}{\GHz}], $F$ is the number of floors traversed by the signal, and $W$ is the number of walls traversed.
If the signal intersects the corner, we consider that it traverses two walls.
Finally, $\mathbb{I}(x)$ equals $1$ if $x$ is true and $0$, otherwise.
Thus $P(A,B)=P_0-PL(d_{A,B})$, where $P_0$ is the transmitting power, and $d_{A,B}$ is the distance between $A$ and $B$.

To find how often location condition \eqref{eq:loc} is met, we look through various positions of APs and the STA and for each set of positions calculate the number $N_{LC}$ of hostile APs, i.e. those APs for which the location condition is satisfied.

For the sake of simplicity and tractability of the results, we consider only regular APs and STAs deployment. 

Specifically, all the APs and the STAs are located at the same height of \SI{1.5}{\m}. We consider 9 possible locations of the APs: in the center of the room, in the middle of each wall, in one of 4 corners 1 m away from each wall. 

Since positions of all STAs except for the considered one do not affect the beacon collision probability, we look over only possible positions of the STA in the considered apartment.
Given integer $m$ and $n$, the sizes of the apartment in meters, we build a $m \times n $ grid with the step of 1 m and offset 0.5 m from the walls. Thus we can analyze how many alien APs can damage the beacon in a particular area of the apartment.

\section{Results and Discussion}
\label{sec:results}

\subsection{Experimental Results}

Let us evaluate how often beacon collisions occur in the described above residential scenario, with numerical parameters listed in Table~\ref{tab:parameters}.
\begin{table}[h]
	\caption{\label{tab:parameters}Scenario Parameters}
	\begin{center}
		\begin{tabular}{| c | c |}
			\hline
			\textbf{Parameter}	& \textbf{Value}							\\ \hline
			Floor height		& \SI{3}{\m}								\\ \hline
			Apartment size		& a) \SI{10}{\m}$\times$\SI{10}{\m}, b) \SI{7}{\m}$\times$\SI{12}{\m}	\\ \hline
			Transmission power	& \SI{18}{\dBm}								\\ \hline
			Receiver threshold	& \SI{-86}{\dBm}							\\ \hline
			Fading parameter	& a) \SI{0}{\dB} b) \SI{3}{\dB} c) \SI{6}{\dB}			\\ \hline
			Carrier frequency	& \SI{2.4}{\GHz}							\\ \hline
			Beacon duration		& \SI{500}{\us}								\\ \hline
			Beacon interval		& \SI{500}{\ms}								\\ \hline
			Carrier frequency	& \SI{2.4}{\GHz}							\\ \hline
			Number of orthogonal channels	& 3								\\ \hline
		\end{tabular}
	\end{center}
\end{table}

\begin{figure}[tb]
	\centering
	\includegraphics[width=0.7\linewidth]{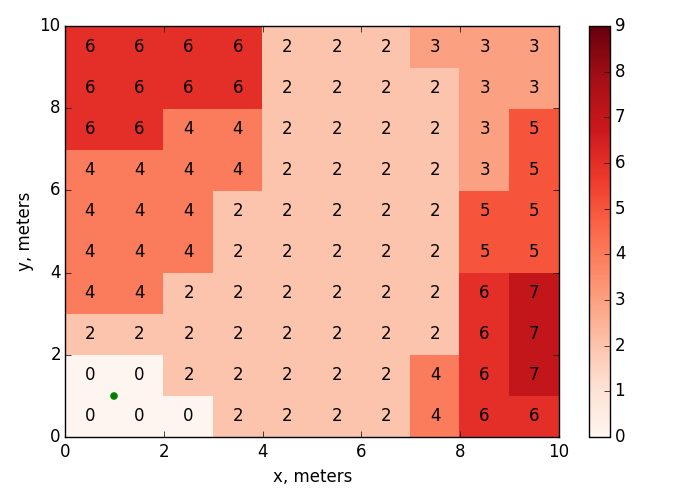}
	\caption{\label{fig:10x10_map_corner_0} The value of $N_{LC}$ for a \SI{10}{\m}$\times$\SI{10}{\m} apartment in the center of the building with APs placed in corner}
\end{figure}

\begin{figure}[tb]
	\centering
	\includegraphics[width=0.7\linewidth]{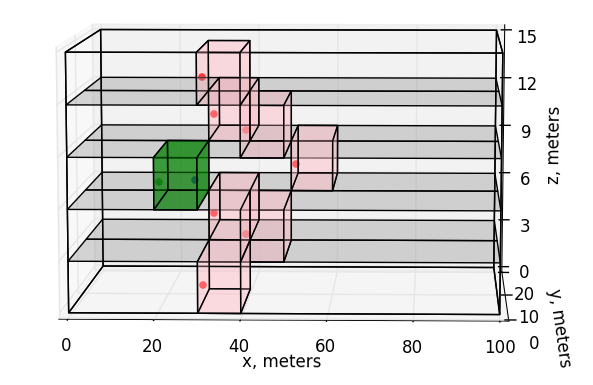}
	\caption{\label{fig:10x10_neighbors_corner_0} The neighbor APs for which LC is met, \SI{10}{\m}$\times$\SI{10}{\m} apartments with APs placed in corner}
\end{figure}

As for the time condition, the probability to meet it for a pair of APs is 0.1\% since the beacon interval is 1000 times longer than the beacon duration.
If there were no clock drifting, having started to collide, the beacons would continue colliding for ages.
However, because of clock drifting, such a continuous collision sometimes ends.
Let us estimate this time.
While the standard \cite{802.11} states that the clock drifting shall be less than 100 ppm, in real devices it is less than 10 ppm.
At the considered time scale, the clock drifting can be considered as a linear process \cite{sivrikaya2004time}.
If the clock drifting of the APs is in opposite directions and it is 10 ppm (relative drifting is 20ppm), beacons of 0.5 ms duration will collide for 25 seconds, which is enough for malfunction to appear.
For small relative drifting, e.g. 1 ppm, the beacons will collide for about 8 minutes. At the same time, even the beacons of two APs do not collide, because of clock drifting sooner or later they will start to collide. Specifically for fast clock drifting (20 ppm), the period between two beacon collision interval is about 7 hours, while for slow clock drifting (1ppm) it is about a week. It means that even low probability of time condition can have dramatic consequences, making users reboot its AP every several days, or even more often.  
  
As for the location condition, let us consider 2.4 GHz band and calculate the number $N_{LC}$ of hostile APs, transmissions of which can prevent the STA from receiving the beacon from $AP_0$ if $\Delta P = 0$.
Fig.~\ref{fig:10x10_map_corner_0} shows a heatmap with such numbers for various STA positions in a central apartment.
If the STA is located in the worst area, there are 7 APs which can prevent it from receiving the desired beacon.
An important fact is that these hostile APs are located in rather far apartments, e.g. in 2 or even 3 hops, see Fig.~\ref{fig:10x10_neighbors_corner_0}.

\begin{figure}[tb]
	\centering
	\includegraphics[width=0.5\linewidth]{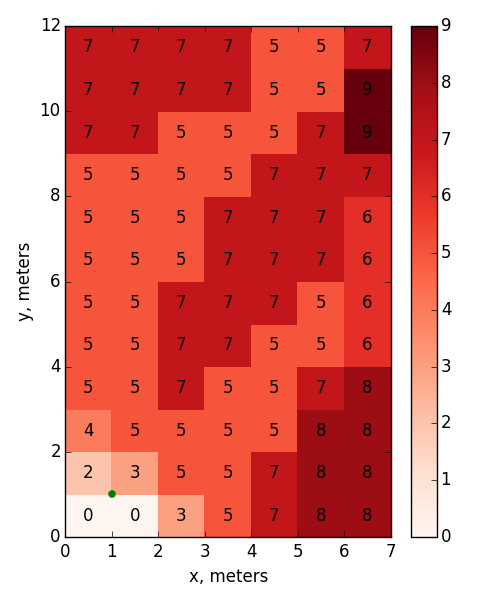}
	\caption{\label{fig:7x12_map_corner_0} The value of $N_{LC}$ for a \SI{7}{\m}$\times$\SI{12}{\m} apartment in the center of the building with APs placed in corner}
\end{figure}

\begin{figure}[tb]
\centering
\includegraphics[width=0.8\linewidth]{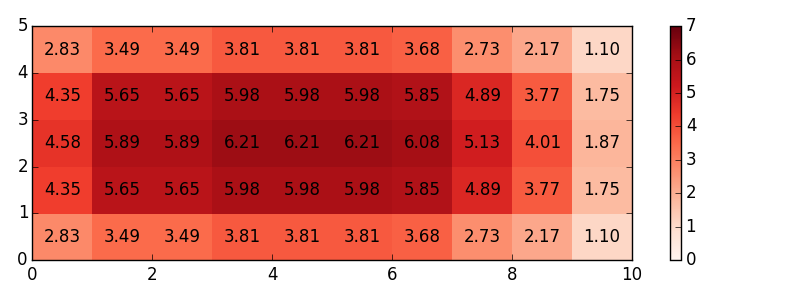}
\includegraphics[width=0.8\linewidth]{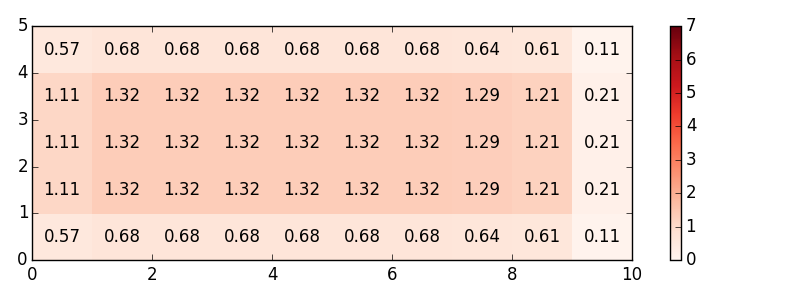}
\caption{\label{fig:7x12_apartments_corner_0} The average number of hostile APs for \SI{7}{\m}$\times$\SI{12}{\m} apartments with APs placed in corner for $\Delta P=0$ (above) and 6 dB (below). The vertical axis corresponds to the floor, the horizontal axis corresponds to a position of an apartment in the row. The considered row contains apartments with APs in the inner /outer corner.}
\end{figure}

The situation becomes even worse when the apartment size is \SI{7}{\m}$\times$\SI{12}{\m} (see Fig.~\ref{fig:7x12_map_corner_0}).
In this case, the number of hostile APs in some areas can reach 9, and there is almost no safety place in an apartment.
The results becomes less crucial if we tighten up the location condition by changing $\Delta P$ from 0 to 6 dB. Specifically, Fig.~\ref{fig:7x12_apartments_corner_0} shows the average number of hostile APs in each apartment for both cases: $\Delta P=0$ and 6 dB.  

\begin{figure}[tb]
	\centering
	\includegraphics[width=0.5\linewidth]{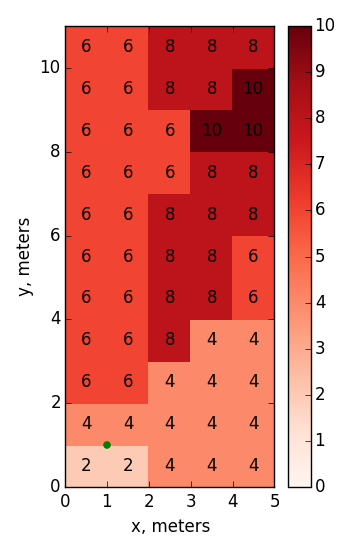}
	\caption{\label{fig:5x11_map_corner_0} The value of $N_{LC}$ for a \SI{5}{\m}$\times$\SI{11}{\m} apartment in the center of the building with APs placed in corner, \SI{5}{\GHz} band}
\end{figure}

\begin{figure}[tb]
	\centering
	\includegraphics[width=\linewidth]{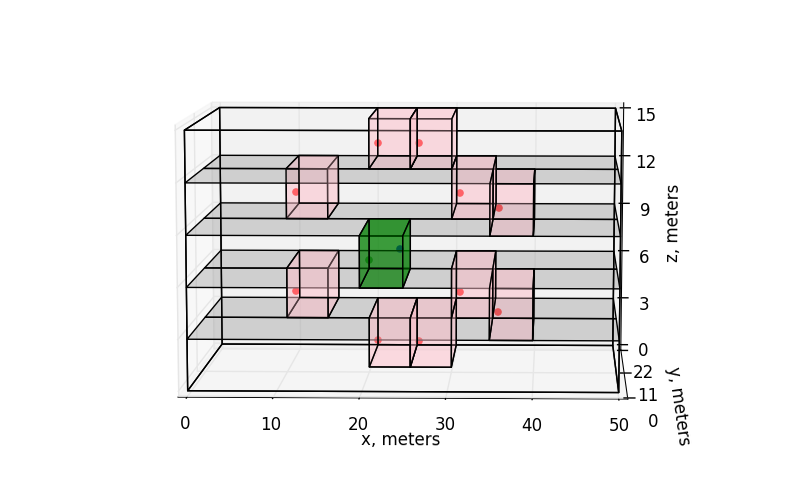}
	\caption{\label{fig:5x11_neighbors_corner_0} The neighbor APs for which LC is met, \SI{5}{\m}$\times$\SI{11}{\m} apartments with APs placed in corner, \SI{5}{\GHz} band}
\end{figure}

\begin{figure}[tb]
\centering
\includegraphics[width=0.8\linewidth]{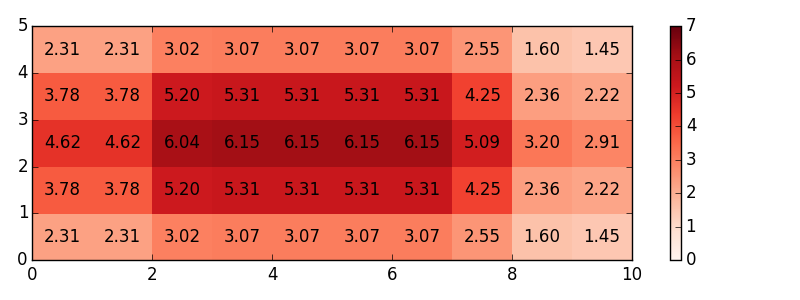}
\includegraphics[width=0.8\linewidth]{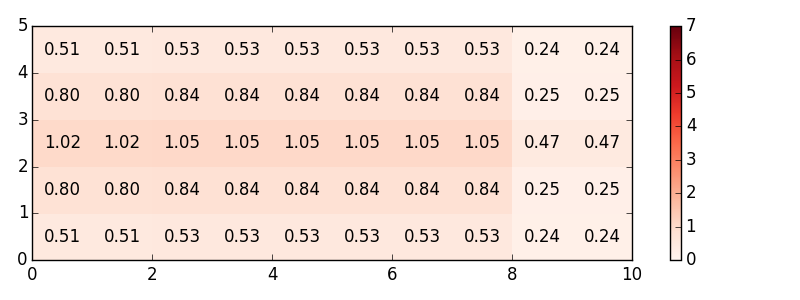}
\caption{\label{fig:5x11_apartments_corner_0} The average number of hostile APs for \SI{5}{\m}$\times$\SI{11}{\m} apartments with APs placed in corner for $\Delta P=0$ (above) and 6 dB (below). \SI{5}{\GHz} band is considered. The vertical axis corresponds to the floor, the horizontal axis corresponds to a position of an apartment in the row. The considered row contains apartments with APs in the inner /outer corner.}
\end{figure}

For \SI{5}{\GHz} band we have obtained similar results.
They are shown in Figures \ref{fig:5x11_map_corner_0}, \ref{fig:5x11_neighbors_corner_0}, \ref{fig:5x11_apartments_corner_0}.
As one can see, the number of hostile APs can reach 10, and even being in range of \SI{1}{\m} from the AP the STA cannot be safe from beacon collisions.

%

\subsection{Discussion}

%
%
As the obtained results have shown, the number of hostile APs can be much greater than the number of orthogonal channels in \SI{2.4}{\GHz} and is comparable to the number of available \SI{20}{\MHz} channels in \SI{5}{\GHz}.
Let us consider several possible solutions which eliminate beacon collisions.

The first approach is to make the beacon interval different for various APs.
Since a single beacon collision itself does not lead to significant problems, choosing beacon intervals with small greatest common devision may avoid aforementioned  malfunctions by preventing continuous beacon collisions. 

Although this approach allows avoiding continuous beacon-to-beacon collisions, it does not protect beacons from collisions with data, sent by hidden APs.
A possible solution is to use a Mesh Beacon Collision Avoidance (MBCA) protocol, adapted to the dense network deployment.
The protocol needs some modification since mesh STAs typically work in same channel, every mesh STA transmits beacons and beacon interval is the same for all the mesh STAs, while in infrastructure networks, beacons are sent only by APs and beacon intervals may differ.
Let us briefly describe the modified version.
AP can include an information element in its beacons, which describes the channel, transmission time, estimated duration and repetition interval of every beacon received by that AP.
Having received such an element from an alien AP, the AP adjusts its beacon transmission time to avoid intersection with that AP.
An AP learns that its beacons collide, if the parameters of its beacons are not included in neighboring APs' beacons.
To avoid deadlock in some corner scenarios, transmissions of beacons are sometimes delayed by a random value.
Finally, neither APs nor STAs can transmit data packets if their transmissions overlap any beacons.

Apart from that, beacon collision can be avoided with the proper use of Dynamic Sensitivity Control (DSC)\cite{afaqui2015evaluation} currently discussed in 802.11ax. DSC allows the STAs to dynamically select their sensitivity level, based on the signal strength come from their own AP.
Increasing this level allows the STAs to ignore low-power frame preambles received from far APs, consider the medium as idle and successfully receive their own beacons.
Thus, if the APs have lower sensitivity level than the STAs, the situation when a STA receives a beacon, but its AP does not receive it becomes impossible.
In other words, using DSC increases $\Delta P$, which decreases the probability that location conditions are met.

%
%
%
%

\section{Conclusion}
\label{sec:conclusion}

In this paper, we have studied the beacon collision problem in dense Wi-Fi networks.
Although the probability that the time condition is met is rather low, the consequences of such rare but continuous collisions are dramatic leading to ``inexplicable'' occasional malfunction, which may cause dissociation from the APs and which makes users continuously reboot their AP to fix the problem.
We have shown that in a canonical residential scenario, location conditions are met for a high number of APs which is much higher than the number of channels in a \SI{2.4}{\GHz} band. The problem is also typical for the \SI{5}{\GHz} band which makes it of high importance for IEEE 802.11ax developers. 

It is not clear yet, which beacon protection mechanism will be finally approved by TGax.
Anyway, until a solution is specified, manufactures should use workarounds in their devices developed for dense deployment.
At least, they should not expect that continuous beacon losses always means that the AP is out of the transmission area of their devices.

\section{Acknowledgments}
The reported study was partially supported by Quantenna Communications and by RFBR (Russian Foundation for Basic Research), research project No. 15-37-70004 mol\_a\_mos.

\ifCLASSOPTIONcaptionsoff
  \newpage
\fi

\bibliographystyle{ieeetr}
\bibliography{biblio.bib}

\end{document}